\documentclass{amsart}
\usepackage{amssymb}
\usepackage{amscd,cite}
\usepackage{psfig}

\newcommand{\sech}{\operatorname{sech}}

\newcommand{\sla}{\operatorname{sl}}

\newcommand{\lp}{\left(}
\newcommand{\rp}{\right)}

\newcommand{\hT}{\hat{T}}

\newcommand{\cP}{\mathcal{P}}
\newcommand{\cE}{\mathcal{E}}

\begin{document}
\title[Quasi-exact solvability beyond the $SL(2)$
algebraization]{Quasi-exact solvability beyond the ${\rm sl}(2)$
algebraization}
\author{David Gomez-Ullate}
\address{ Dep. Matem\`atica Aplicada I, Universitat Polit\`ecnica de Catalunya, ETSEIB, Av. Diagonal 647, 08028 Barcelona, Spain. }
\author{ Niky Kamran }
\address{Department of Mathematics and Statistics, McGill University
Montreal, QC, H3A 2K6, Canada}
\author{Robert Milson}
\address{Department of Mathematics and Statistics, Dalhousie University, Halifax, NS, B3H 3J5, Canada}
\begin{abstract}
We present evidence to suggest that the study of one dimensional
quasi-exactly solvable (QES) models in quantum mechanics should be
extended beyond the usual $\sla(2)$ approach. The motivation is
twofold: We first show that certain quasi-exactly solvable
potentials constructed with the $\sla(2)$ Lie algebraic method
allow for a new larger portion of the spectrum to be obtained
algebraically. This is done via another algebraization in which
the algebraic hamiltonian cannot be expressed as a polynomial in
the generators of $\sla(2)$. We then show an example of a new
quasi-exactly solvable potential which cannot be obtained within
the Lie-algebraic approach.
\end{abstract}

\maketitle

\section{Introduction}

Lie-algebraic and Lie group theoretic methods have played a
significant role in finding exact solutions to the Schr\"odinger
equation in quantum mechanics. In the classical applications, the
Lie group appears as a symmetry group of the Hamiltonian operator,
and the associated representation theory provides an algebraic
means for computing the spectrum. Of particular importance are the
exactly solvable problems, such as the harmonic oscillator or the
hydrogen atom, whose point spectrum can be completely determined
using purely algebraic methods. The concept of a {\em spectrum
generating algebra} dates back to a paper by Goshen and Lipkin
\cite{GL}, and was later rediscovered in the context of high
energy physics by two different groups \cite{BB,DGN}. The study of
spectrum generating algebras received much impetus in the
subsequent years (see the review \cite{BN}) and it was soon
applied also in the field of molecular dynamics by Iachello,
Levine, Alhassid, G\"ursey and their collaborators (see the book
\cite{IL} for a survey of theory and applications). Most of the
applications of spectrum generating algebras concerned {\em
exactly solvable} Hamiltonians whose spectrum could be completely
determined by algebraic means. An intermediate class of spectral
problems are those for which only a finite part of the point
spectrum can be calculated by algebraic methods, but possibly not
the whole spectrum. The usual example of these type of problems is
the sextic oscillator in which the potential is an even polynomial
of degree six whose coefficients depend on a parameter $n$. For
each positive integer value of $n$ the Hamiltonian is shown to
preserve an $(n+1)$-dimensional subspace of $L_2(\mathbb R)$. It
is clear from Sturm-Liouville theory that the Hamiltonian with the
sextic potential has an infinite number of bound states, but only
$n+1$ of them belong to the so called {\em algebraic} sector. It
was soon realized that the sextic oscillator was just the first
example of a large class of systems having this property, and the
works of  Shifman, Turbiner, Ushveridze \cite{ST,T,U1} and their
collaborators initiated the study of the mathematical properties
and physical applications of this new class of spectral problems,
which they named {\em quasi-exactly solvable}.
However, even in the simplest case of one spatial dimension, there
is no way to ascertain whether a given potential is quasi-exactly
solvable or not, so two methods were proposed to construct large
families of quasi-exactly solvable problems: the Lie-algebraic
method \cite{T} and the analytic method \cite{U2}.

The idea underlying the Lie-algebraic method is to use results
 from the representation theory of Lie algebras to ensure that a
 given Hamiltonian preserves a certain finite dimensional
 subspace of functions. In one dimension, the only algebra of
 first order differential operators with finite dimensional is
 $\sla(2)$, whose generators:
\begin{eqnarray}
\nonumber
  J_n^+ &=&z^2 D_{z}-n z,\\ \label{eq:sl2generators}
  J_n^0 &=& z D_{z} - \frac{n}{2},\\
  J_n^- &=&   D_{z}\nonumber
\end{eqnarray}
preserve the $(n+1)$-dimensional linear space of polynomials in
the $z$ variable of degree less than or equal to $n$:
\begin{equation}\label{Pnz}
\mathcal P_n(z) = \langle 1,z,z^2,\dots,z^n \rangle\,.
\end{equation}
The most general second order differential operator that preserves
$\mathcal P_n(z)$ can be written as a quadratic combination of the
generators (\ref{eq:sl2generators}) of $\sla(2)$
\begin{equation}
\overline H=\sum c_{a,b} J_n^a J_n^b + \sum c_{a} J_n^a +c^*,
\end{equation}
where the indexes run over the values $a,b=\pm,0$. The operator
$\overline H$ is said to be {\em Lie algebraic} and  $\sla(2)$ is
often referred to as a hidden symmetry algebra. The operator
$\overline H$ does not have in general the form of a Schr\"odinger
operator, but it can be transformed into a Schr\"odinger operator
by a change of variables and a conjugation by a non-vanishing
function. Such a transformation always exists in one spatial
dimension, but in more dimensions the equivalence problem remains
open \cite{M,GKO91}. This has been a long standing obstacle to
classify multidimensional quasi-exactly solvable Hamiltonians,
where only a few families are known \cite{GGR00,GGR01}.

One important fact lies at the core of the Lie-algebraic method:
it is ensured by Burnside's classical theorem that every
differential operator which leaves the space $\cP_n(z)$ invariant
belongs to the enveloping algebra $\mathcal U(\sla(2))$, since
$\mathcal P_n$ is an irreducible module for the $\sla(2)$ action.
This is probably the reason for the relative success of the
 Lie-algebraic constructions in the context of quasi-exact
solvability, up to the point that the terms {\em Lie-algebraic}
and {\em quasi-exactly solvable} are often used as synonims in the
literature.

However, there are many other finite dimensional polynomial spaces
 which are not irreducible modules for the $\sla(2)$ action, and  in
 these cases there might be non-Lie algebraic differential operators which leave the space invariant.
The first exploration of this type was done by Post and Turbiner
\cite{PT}, who classified all second order differential operators
which preserve a polynomial space generated by monomials. In their
work they did not use Lie algebras but other considerations based
on the grading of the operators and basis elements. It was later
realized that these differential operators can also be transformed
into Schr\"odinger form thereby providing new examples of
quasi-exactly solvable operators which are not Lie algebraic. In
this {\em direct approach} to quasi-exact solvability \cite{GKM3}
more general polynomial spaces are considered and the set of
differential operators that preserve them are investigated without
any reference to Lie algebras. It was later shown that exactly
solvable potentials exist which are not Lie algebraic, but can be
obtained from a Lie algebraic potentials via a state-adding
Darboux transformation \cite{GKM1}. Further research along this
line suggests to regard a Darboux or SUSY transformation not just
as a transformation on the potential and eigenfunctions, but as an
algebraic deformation of an infinite polynomial flag \cite{GKM2}.
Non Lie-algebraic potentials  appear also  in the recent work of
Gonz\'alez-L\'opez and Tanaka  in the context of supersymmetry
\cite{GT05}.

In contrast with the Lie algebraic method, the analytic method has
received less attention over the last decades. However recent
results suggest that it is a more suitable approach to encompass
this new type of quasi-exactly solvable systems which do not fit
in the Lie algebraic scheme,\cite{GKMnew}.

In this paper we would like to illustrate the relevance of
studying quasi-exact solvability beyond the Lie algebraic approach
by providing two suitably chosen examples: in Section
\ref{sec:more} we show how more energy levels can be obtained from
a Lie algebraic potential which cannot be obtained in the
$\sla(2)$ approach. In Section \ref{sec:noLie} we show an example
of a potential which is quasi-exactly solvable  but not Lie
algebraic.

\section{Calculating more algebraic energy levels of a Lie algebraic
potential}\label{sec:more}

In this Section we expose the first reason to study quasi-exactly
solvable problems beyond the Lie-algebraic approach. We present a
potential which is known to admit an $\sla(2)$ algebraization, and
therefore allows for some of its energy levels to be calculated
algebraically, all belonging to the even sector. For this same
potential, we show that a different algebraization exists in which
the algebraic Hamiltonian cannot be expressed as an element of the
enveloping algebra of $\sla(2)$. This new algebraization allows to
calculate all the previous levels, plus some extra ones.

Consider the following Schr\"odinger operator
\begin{equation}\label{eq:sch}
H=-D_{xx} + 2A^2  \cosh(2x) + 4A\,n \cosh(x)-\frac{1}{2}
\sech^2(x/2),
\end{equation}
where $A<0$ is a real parameter and where $n$ is a positive
integer.  This potential is known to be Lie-algebraic and it
appears for instance in the classification performed by
Gonz\'alez-L\'opez, Kamran and Olver in \cite{GKO93}. The
$\sla(2)$ algebraization of this potential is achieved by the
transformation
\[T=\mu(x)^{-1}H \mu(x),\] with the following choice of gauge factor and
change of coordinate
\begin{eqnarray}
\mu(x)&=&e^{2A\cosh(x)} \sech(x/2),\label{mu1}\\
  z &=& -\sinh^2(x/2).\label{change1}
\end{eqnarray}
Up to an additive constant, the transformed operator $T$ becomes
\begin{equation}\label{eq:T}
  T=z(1-z)\, D_{zz} +\big( 8Az(z-1) +\frac{1}{2}\big) \, D_z
  -8Anz
\end{equation}
which is easily seen to preserve the space $\mathcal P_n(z)$
defined in (\ref{Pnz}). The explicit quadratic combination of $T$
in terms of $\sla(2)$ generators is (again up to an additive
constant)
\begin{equation}
  T= -(J_n^0)^2 +\frac{1}{2}\, \{ J_n^0, J_n^-\} +8A\, J_n^+ +(1-8A-n)\, J_n^0
  +\frac{n}{2}\, J_n^-,
\end{equation}
As a consequence of this $\sla(2)$ algebraization we can calculate
$\dim \cP_n=n+1$ energy levels of the Hamiltonian, by
diagonalizing the corresponding matrix of the restricted action of
$T$ to $\cP_n$. The $(n+1)$ algebraic eigenfunctions  have the
form
\begin{equation}
\psi_{2k}(x)= \mu(x)\, p_k(-\sinh^2(x/2))\,,\qquad k=0,\dots,n
\end{equation}
where $p_k$ is one of the $(n+1)$ polynomial eigenfunctions that
$T$ is ensured to have. All the eigenfunctions obtained via the
Lie algebraic method correspond to the even sector.

Up to this point, all these results are well known. However,
 the {\em same} Hamiltonian remarkably admits a {\em different}
algebraization characterized by the following gauge factor and
change of variables:
\begin{eqnarray}
\hat\mu(x)&=&{\rm e}^{2A\cosh(x)} \,{\rm e}^{-nx}\,\sech(x/2), \label{mu2}\\
  w &=& {\rm e}^x.\label{change2}
\end{eqnarray}
The transformed operator $\hat T=\hat\mu(x)^{-1} H \,\hat\mu(x)$
becomes
\begin{equation}
  \hT = -w^2 D_{ww} +\lp -2A\,w^2 + 2\,n\,w+2(A-1)+\frac{2}{1+w}\rp
  D_{w} +4A\,n\,w   + \frac{2\,n}{1+w},
\end{equation}
where an additive constant has been dropped. It is straightforward
to realise that due to the rational coefficients in the previous
expression, the operator $\hat T$ does not belong to the
enveloping algebra of $\sla(2)$. Nevertheless it preserves a
$2n$-dimensional subspace generated by polynomials in the variable
$w$ that we shall denote as $\cE^{(n,-1)}_{2n}(w)$ and define as
\begin{equation}
\label{eq:emodtest}
 \cE^{(n,-1)}_{2n} = \{ p\in\cP_{2n} \colon  p'(-1)+ np(-1) = 0 \}
\end{equation}
This polynomial subspace is an example of an {\em exceptional
polynomial module}, the name exceptional being due to the fact
that the space of second order operators that leave it invariant
have a rich structure \cite{GKM3,GKMnew}. An explicit basis of
$\cE^{(n,-1)}_{2n}(w)$ can be given in the following manner
\begin{equation}
\cE^{(n,-1)}_{2n}(w)=\left\langle
1-n(w+1),(w+1)^2,(w+1)^3,\dots,(w+1)^{2n} \right\rangle
\end{equation}
This second algebraization allows us to compute $2n$ algebraic
eigenfunctions of the Schr\"odinger operator (\ref{eq:sch}), which
will be of the form
\begin{equation}
\psi_{k}(x)= \hat\mu(x)\, \hat p_k({\rm e}^x)\,,
\end{equation}
where $\hat p_k(w)$ is one of the $2n$ polynomial eigenfunctions
that $\hat T$ is ensured to have. We observe that these
eigenfunctions can be both odd and even, as opposed to the
$\sla(2)$ algebraic ones, which were only even.

\begin{figure}[h]
\begin{center}
\begin{tabular}{c}
\psfig{figure=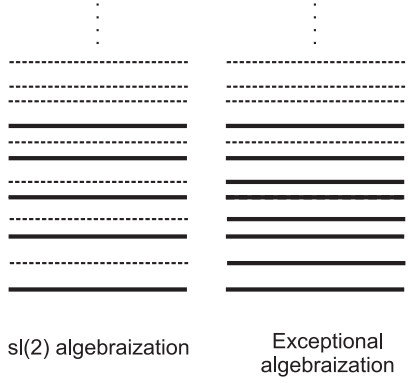,width=2.2in}
\end{tabular}
\end{center}
\caption{The algebraic energy sector coming from both
algebraizations of potential (\ref{eq:sch}) with
$n=4$.}\label{fig0}
\end{figure}

We thus obtain $n+1$ algebraic eigenfunctions via the $\sla(2)$
algebraization and $2n$ algebraic eigenfunctions via the
exceptional module algebraization. The situation is depicted
schematically in Figure \ref{fig0}.
It is maybe worth to explore the
 relation between the two different algebraizations. It
is known in the theory of Lie algebraic problems that under linear
fractional transformations the Lie-algebraic character is
preserved \cite{GKO93}. More specifically, a change of variable by
a linear fractional transformation,
\[
z=\frac{\alpha {\tilde z} +\beta}{\gamma {\tilde z} +
\delta}\,,\qquad \alpha\delta-\beta\gamma=1,
\]
together with an appropriate gauge transformation preserve the
space $\cP_n$ since
\begin{equation}\label{2multrep}
p(z)\in\cP_n(z) \mapsto \tilde p({\tilde z}) = (\gamma {\tilde
z}+\delta)^n p \left(\frac{\alpha {\tilde z}+\beta}{\gamma {\tilde
z} + \delta}\right)\in \cP_n({\tilde z})\,.
\end{equation}
Moreover, if $T$ is Lie-algebraic and preserves $\cP_n(z)$, then
\begin{eqnarray}
\tilde T &=& (-\gamma z + \alpha)^{-n} \, T\,(-\gamma z
+\alpha)^{n}
\end{eqnarray}
is also Lie-algebraic and preserves $\cP_n({\tilde z})$. We see
thus that projective transformations do not change the $\sla(2)$
character, in fact they amount to a linear transformation of the
generators of $\sla(2)$ \cite{GKO93}. However, from expressions
(\ref{mu1}),(\ref{change1}),(\ref{mu2}) and (\ref{change2}) it
follows that the relation between the two algebraic operators $T$
and $\hat T$ in the previous example is given by
\begin{eqnarray}
  z &=& -\frac{(w-1)^2}{4w},\\
  T &=&  w^{-n}\, \hat T\, w^{n},
\end{eqnarray}
which is not a projective transformation.

We have performed an explicit computation to show a few of the
algebraic eigenfunctions of the Hamiltonian (\ref{eq:sch}). For
illustrative purposes it suffices to fix the integer parameter $n$
to some low value. For the values $n=3$ and $A=-0.4$ the potential
(\ref{eq:sch}) has the shape of a double-well potential shown in
Figure \ref{fig1}.

\subsection{$\sla(2)$ algebraization}

\begin{figure}[h]
\begin{center}
\begin{tabular}{cc}
\begin{tabular}{c}
\psfig{figure=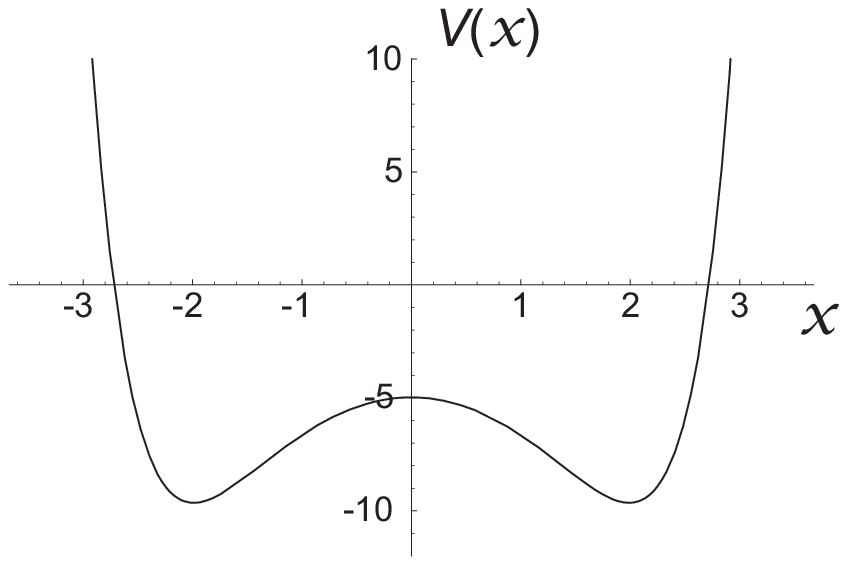,width=2.2in}\\{ \scriptsize{Potential
$V(x)$}}
\end{tabular} &
\begin{tabular}{c}
\psfig{figure=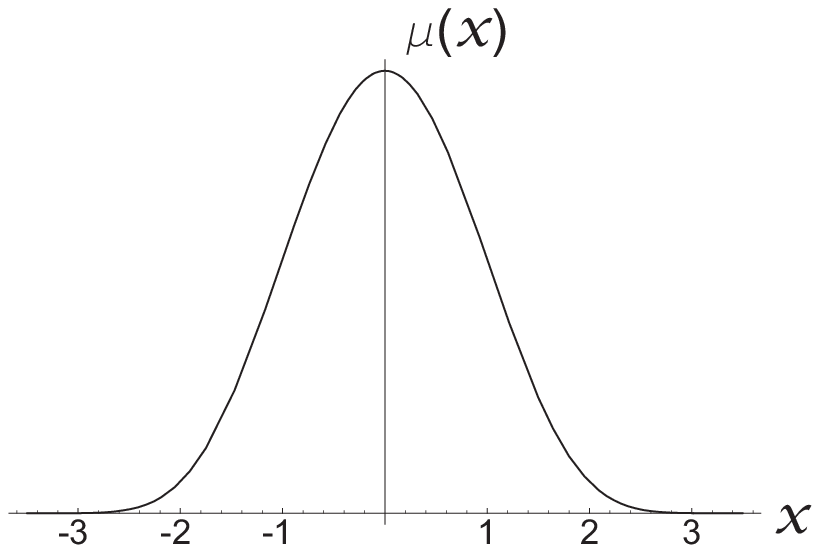,width=2.2in}\\ {\scriptsize{ Gauge factor
$\mu(x)$} }
\end{tabular}\\\\
\begin{tabular}{c}
\psfig{figure=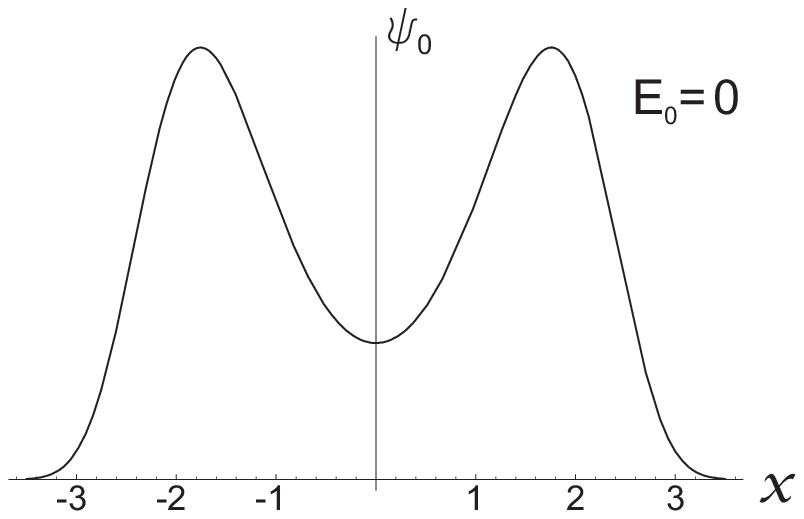,width=2.2in}\\ {\scriptsize{Ground state
$\psi_0(x)$}}
\end{tabular} &
\begin{tabular}{c}
\psfig{figure=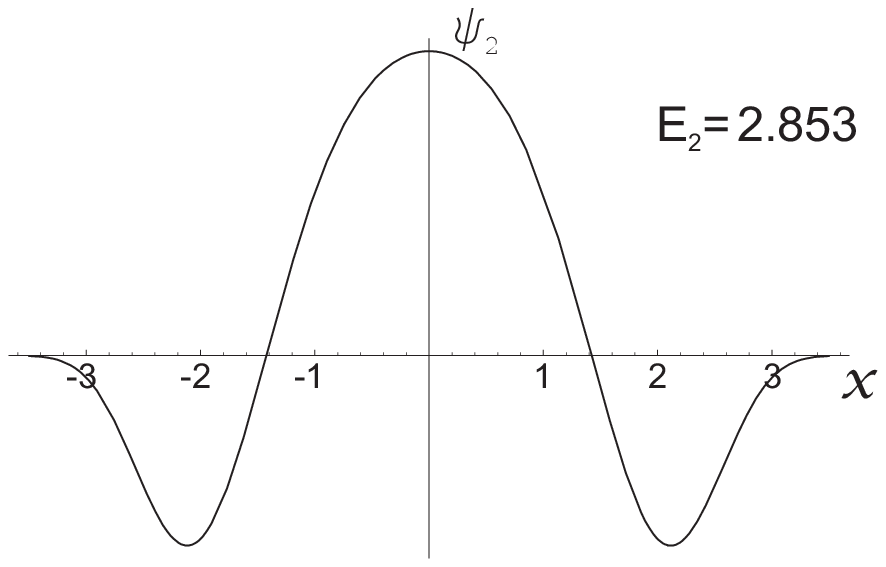,width=2.2in}\\ {\scriptsize{Second excited
state $\psi_2(x)$}}
\end{tabular}
\\\\
\begin{tabular}{c}
\psfig{figure=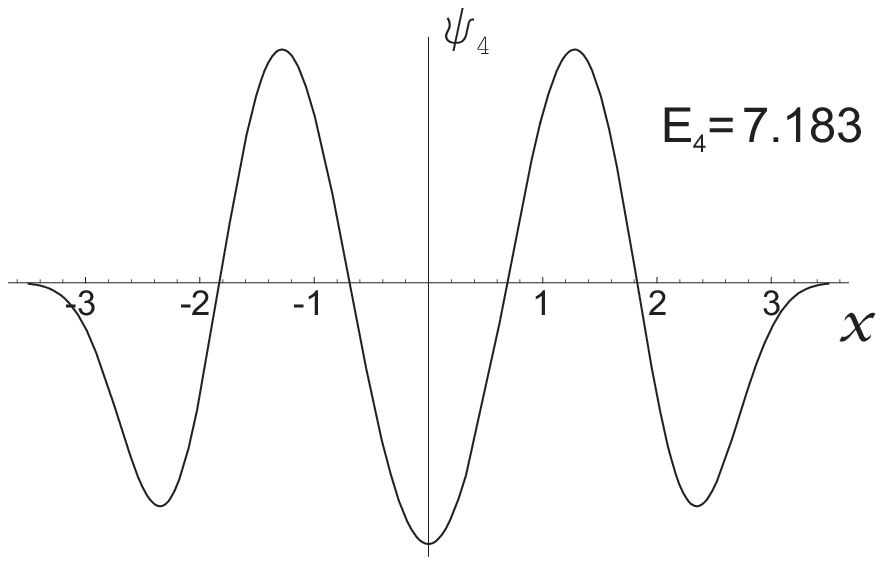,width=2.2in}\\
{\scriptsize{Fourth excited state $\psi_4(x)$}}
\end{tabular} &
\begin{tabular}{c}
\psfig{figure=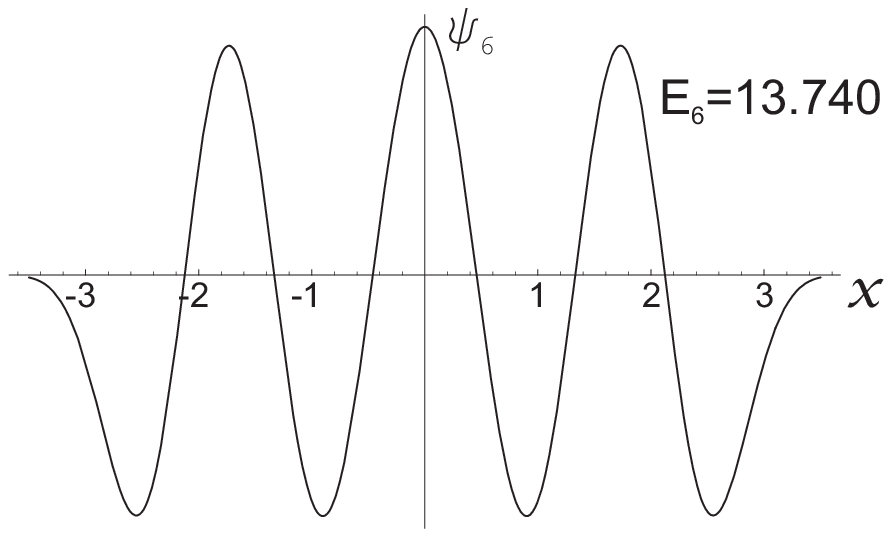,width=2.2in}\\ {\scriptsize{Sixth excited
state $\psi_6(x)$}}
\end{tabular}

\end{tabular}
\end{center}
\caption{The four algebraic eigenfunctions obtained through
$\sla(2)$ algebraization of potential (\ref{eq:sch}) with $n=3$
and $A=-0.4$ }\label{fig1}
\end{figure}

For $n=3$ the operator $T$ in (\ref{eq:T}) satisfies
\[T\, \cP_3(z)\subset\cP_3(z)\]
which allows to compute four algebraic eigenfunctions via the
$\sla(2)$ algebraization. The action of $T$ relative to the
canonical basis of $\cP_3$ is given by the following $4\times4$
matrix:
\[
T\Big|_{\,\cP_3}={\footnotesize\left(%
\begin{array}{cccc}
  0 & -24A & 0 & 0 \\
  1/2 & -8A & -16A & 0 \\
  0 & 3 & -2-16A & -8A \\
  0 & 0 & 15/2 & -6-24A \\
\end{array}%
\right)}.
\]
\vskip0.1cm \noindent For $A=-0.4$ we have calculated four
algebraic eigenfunctions of (\ref{eq:sch}) which have the form
%
\begin{equation}
\psi_{2k}(x) = \mu(x)\,p_k(\sinh^2 {\footnotesize
\frac{x}2)},\qquad k=0,1,2,3,
\end{equation}
where $p_k\in\cP_3(z)$ is one of the four eigen-polynomials of $T$
and the gauge factor $\mu(x)$ is given by (\ref{mu1}). These four
eigenfunctions have been plotted in Figure \ref{fig1} along with
their corresponding energies, which have been normalized so that
the energy of the ground state is zero.

\subsection{Exceptional module algebraization}
\begin{figure}[h]
\begin{center}
\begin{tabular}{cc}
\begin{tabular}{c}
\psfig{figure=potential.eps,width=2.2in}\\{ \scriptsize{Potential
$V(x)$}}
\end{tabular} &
\begin{tabular}{c}
\psfig{figure=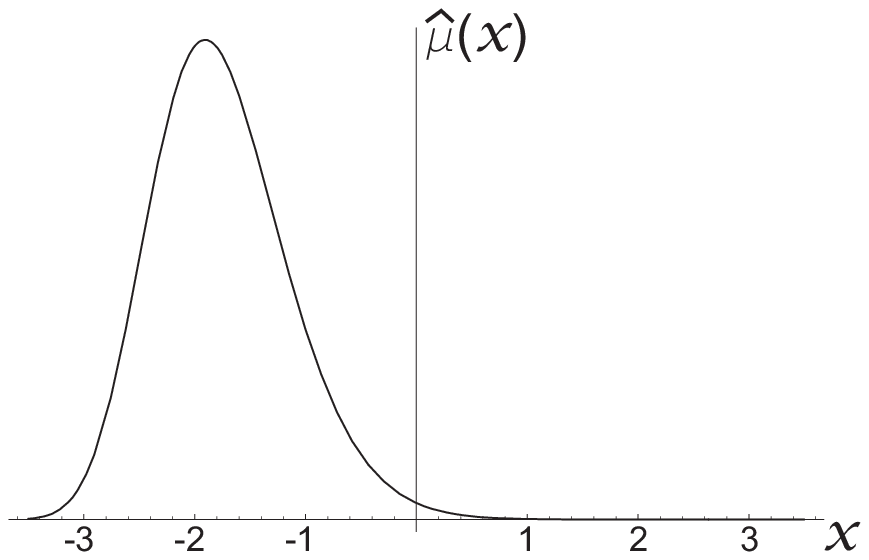,width=2.2in}\\ {\scriptsize{ Gauge factor
$\hat\mu(x)$} }
\end{tabular}\\\\
\begin{tabular}{c}
\psfig{figure=psi0.eps,width=2.2in}\\ {\scriptsize{Ground state
$\psi_0(x)$}}
\end{tabular} &
\begin{tabular}{c}
\psfig{figure=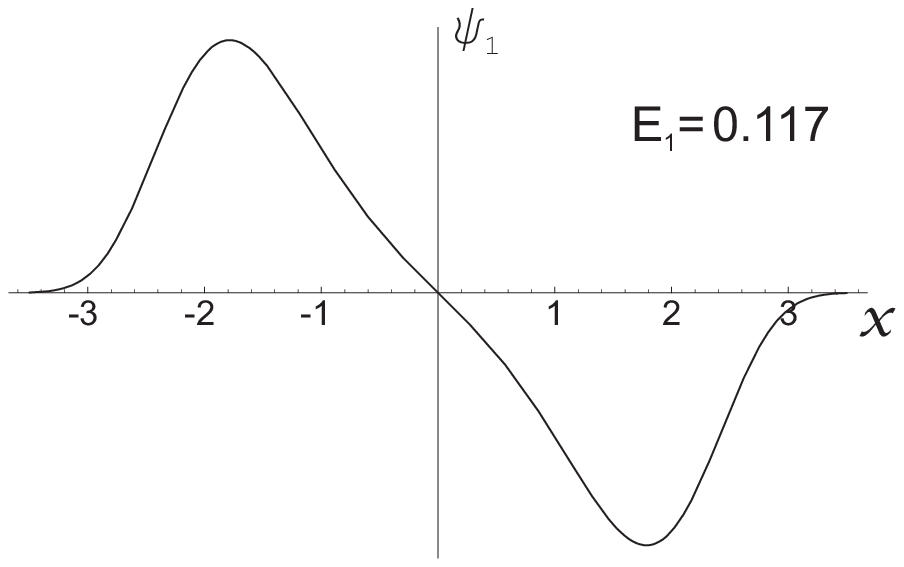,width=2.2in}\\ {\scriptsize{First excited
state $\psi_1(x)$}}
\end{tabular}
\\\\
\begin{tabular}{c}
\psfig{figure=psi2.eps,width=2.2in}\\
{\scriptsize{Second excited state $\psi_2(x)$}}
\end{tabular} &
\begin{tabular}{c}
\psfig{figure=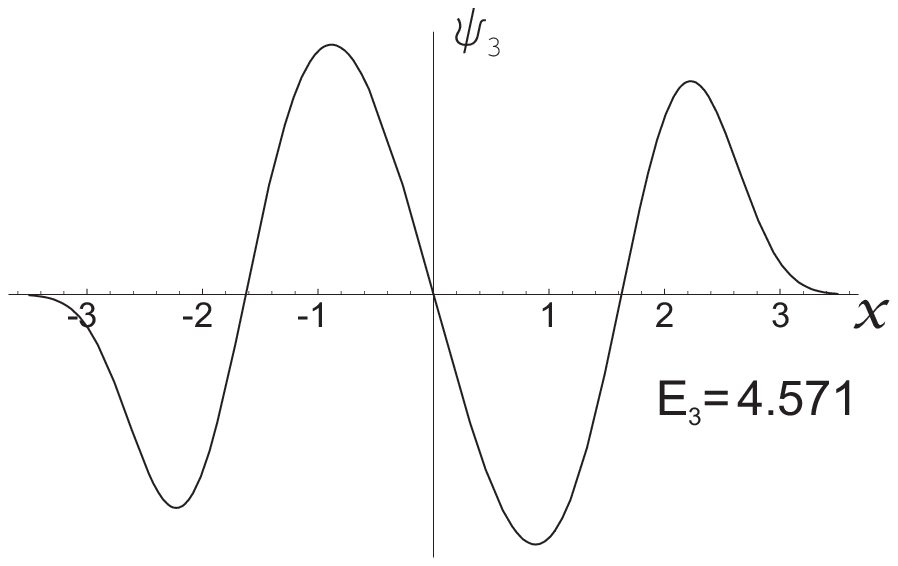,width=2.2in}\\ {\scriptsize{Third excited
state $\psi_3(x)$}}
\end{tabular}
\\\\
\begin{tabular}{c}
\psfig{figure=psi4.eps,width=2.2in}\\ {\scriptsize{Fourth excited
state $\psi_4(x)$}}
\end{tabular} &
\begin{tabular}{c}
\psfig{figure=psi6.eps,width=2.2in}\\ {\scriptsize{Sixth excited
state $\psi_6(x)$}}
\end{tabular}

\end{tabular}
\end{center}
\caption{The six algebraic eigenfunctions obtained through the 
exceptional module algebraization of potential (\ref{eq:sch}) with
$n=3$ and $A=-0.4$ }\label{fig2}
\end{figure}

From the second algebraization we know that the Schr\"odinger
operator $H$ is conjugate via the change of variables
(\ref{change2}) and gauge factor (\ref{mu2}) to the operator $\hat
T$ which preserves an exceptional polynomial module. For $n=3$
this space is spanned by
\begin{equation}
\cE^{(3,-1)}_{6}(w)=\left\langle
-2-3w,(w+1)^2,(w+1)^3,(w+1)^4,(w+1)^5,(w+1)^{6} \right\rangle,
\end{equation}
and the matrix of the action of $\hat T$ relative to this basis is
\vskip0.1cm
\[
\hat T\Big|_{\,\cE^{(3,-1)}_{6}}={\footnotesize\left(%
\begin{array}{cccccc}
  6-12A & -30A & 0 & 0 & 0 & 0 \\
  2 & 10-4A & 8A & 0 & 0 & 0 \\
  0 & -6 & 12 & 6A & 0 & 0 \\
  0 & -4 & -2 & 12+4A & 4A & 0 \\
  0 & 0 & -10& 6 & 10+8A & 2A \\
  0 & 0 & 0 & -18 & 18 & 6+12A \\
\end{array}%
\right)}\,.
\]
\vskip0.1cm \noindent For the same value $A=-0.4$ we have
calculated six algebraic eigenfunctions of the Hamiltonian
(\ref{eq:sch}) which have the form
\begin{equation}
\psi_{k}(x) = \hat\mu(x)\,\hat p_k({\rm e}^x),\qquad
k=0,1,2,3,4,6.
\end{equation}
where $\hat p_k\in \cE^{(3,-1)}_{6}$ is one of the six
eigen-polynomials of $\hat T$ and the gauge factor $\hat\mu(x)$ is
given by (\ref{mu2}). These six algebraic eigenfunctions along
with their energies are shown in Figure \ref{fig2}.  The first
thing to note is that we obtain all the even eigenfunctions from
the $\sla(2)$ algebraization plus two extra odd eigenfunctions
corresponding to the first and third excited states. However, the
eigenfunction corresponding to the fifth excited state is not
present in the algebraic sector. For arbitrary $n$ it seems that
there is always a gap in the algebraic sector just below the
highest energy in the exceptional module algebraization,
\cite{GKM3}. Although all evidences show that this is true in the
general case, there is yet no proof of this result.
%

\section{A non-Lie algebraic quasi-exactly solvable
potential}\label{sec:noLie}

In the previous section we have seen how the exceptional module
algebraization can provide more energy levels from a Lie-algebraic
potential. In this Section we will show that some Schr\"odinger
operators only admit an exceptional module algebraization and not
the traditional $\sla(2)$ one. We can therefore construct new
quasi-exactly solvable potentials on the line, which are not in
the classifications of QES potentials in \cite{GKO93,T}, since
those classifications deal only with potentials that admit an
$\sla(2)$ algebraization. In this Section we show one first simple
example of this phenomenon, postponing the full classification of
these new potentials to a future publication,\cite{GKMnew}.

Consider the following Schr\"odinger operator:
\begin{equation}
H=-D_{xx} + V(x)
\end{equation}
with potential
\begin{equation}\label{eq:sextic}
V(x)= A^2 x^6 + 2 ABx^4 + \big(B^2+ [(-1)^p-4n]A\big) x^2 + 4
\frac{x^2-1}{(x^2+1)^2}.
\end{equation}
If the last rational term were absent, this potential would be the
well known quasi-exactly solvable sextic potential discussed for
instance in \cite{ST}. The potential is always even, so its
eigenfunctions will have well defined parity.  However, as it
happens with the sextic, only even or odd eigenfunctions but not
both appear in the algebraic sector. This does not exclude in
principle that other algebraizations exist in which both even and
odd eigenfunctions are obtained. In fact, the choice of $p=0$
gives a potential with algebraic even eigenfunctions while $p=1$
corresponds to potentials with odd algebraic eigenfunctions. For
arbitrary values of $A$ and $B$ the Hamiltonian does not preserve
any finite dimensional subspace, but for the following values:
\begin{eqnarray}
A&=&\frac{a}{n-a}\left( a-\frac{(-1)^p}2\right),\\
 B&=& \frac{a}{n-a}\left( 3a-2n+(-1)^p \big(\frac{n}{2a}-1\big)
 \right),
\end{eqnarray}
where $a$ is an arbitrary real parameter, the Hamiltonian $H$
admits an exceptional module algebraization. More specificallly,
the transformation  $ T=\mu(x)^{-1} H \,\mu(x)$ with
\begin{eqnarray}\label{musextic}
\mu(x)&=&\frac{x^p}{x^2+1}\,{\rm exp}\big(-\frac{A}4 x^4-\frac{B}2 x^2\big),\quad p=\{0,1\}\\
w&=&x^2
\end{eqnarray}
transforms $H$ into the algebraic operator $T$ given by
\[
T(w) = 4 \big(J_4 +  J_5 -A J_6\big)
\]
up to an additive constant. The operators $J_4$, $J_5$ and $J_6$
are given by
\begin{eqnarray}
J_4&=&(w-1) D_{ww}+\big(a(w-1)-1\big)D_w,\\
J_5 &=& D_{ww} +2 \Big( a-\frac{1}{w-1}\Big)D_w-\frac{2a}{w-1},\\
J_6 &=& (w-1)\Big( w-1-\frac{n}a\Big)D_w -n(w-1),
\end{eqnarray}
and every one of them leaves invariant the exceptional module
$\cE_n^{(a,1)}(w)$ given by
\begin{equation}\label{eq:basis}
\cE_n^{(a,1)}(w)=\big\langle
1-a(w-1),(w-1)^2,(w-1)^3,\dots,(w-1)^n\big\rangle
\end{equation}
The sextic potential (\ref{eq:sextic}) has eigenfunctions in
$L_2(\mathbb R)$ provided that $A<0$, or $A=0$ and $B<0$. This
last case corresponds in fact to an exactly solvable potential,
which can be obtained by a SUSY transformation from the harmonic
oscillator \cite{GKM1}. The algebraic operator preserves in this
last case an infinite flag of exceptional polynomial subspaces.

We show some explicit eigenfunctions of this potential
corresponding to the even sector by setting $n=4$ and $p=0$. The
matrix of the action of $T$ with respect to the basis
(\ref{eq:basis}) of $\cE_4^{(a,1)}(w)$ is

\[
 T\Big|_{\,\cE^{(a,1)}_{4}}={\footnotesize\left(%
\begin{array}{cccc}
  \frac{-15a^2-a+6}{a-4} & \frac{6a^2(1-2a)}{a-4} & 0 & 0  \\
  -8 & \frac{-19a^2+47a-10}{a-4} & \frac{4a(2a-1)}{a-4} & 0 \\
  0 & 4(4a-3) & \frac{-23 a^2+79a-18}{a-4} & \frac{2a(2a-1)}{a-4}\\
  0 & 16 & 8(3a-4) & \frac{-27a^2+111a-26}{a-4}  \\
\end{array}%
\right)}\,.
\]

\begin{figure}[h]
\begin{center}
\begin{tabular}{cc}
\begin{tabular}{c}
\psfig{figure=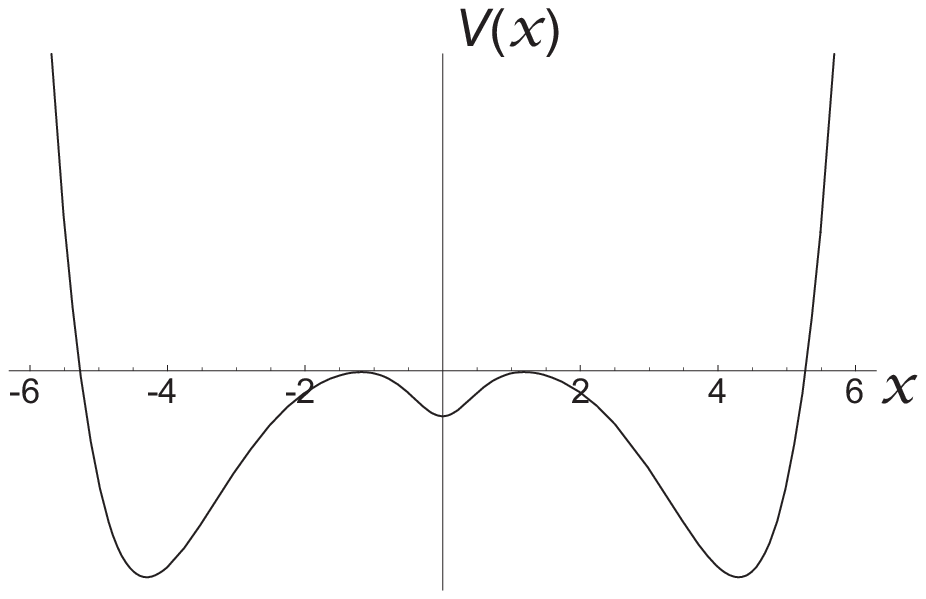,width=2.2in}\\{ \scriptsize{Potential
$V(x)$}}
\end{tabular} &
\begin{tabular}{c}
\psfig{figure=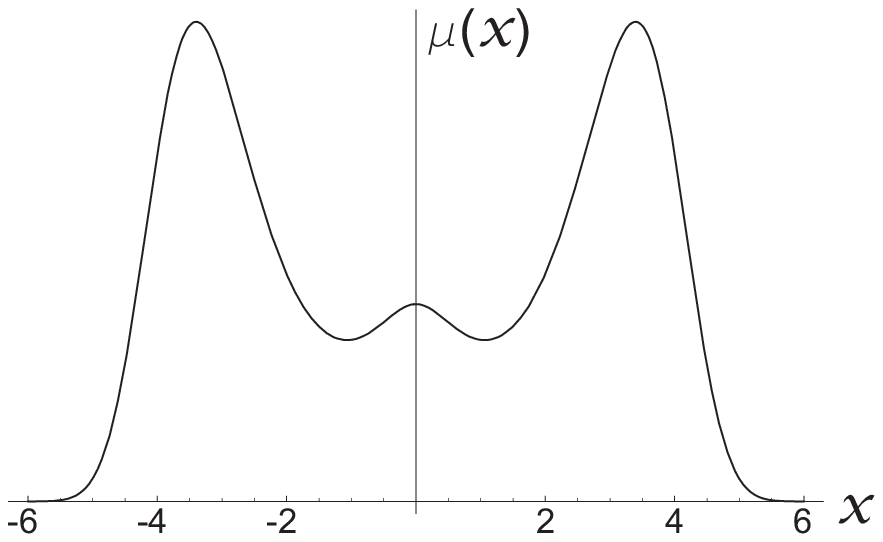,width=2.2in}\\ {\scriptsize{ Gauge
factor $\mu(x)$} }
\end{tabular}\\\\
\begin{tabular}{c}
\psfig{figure=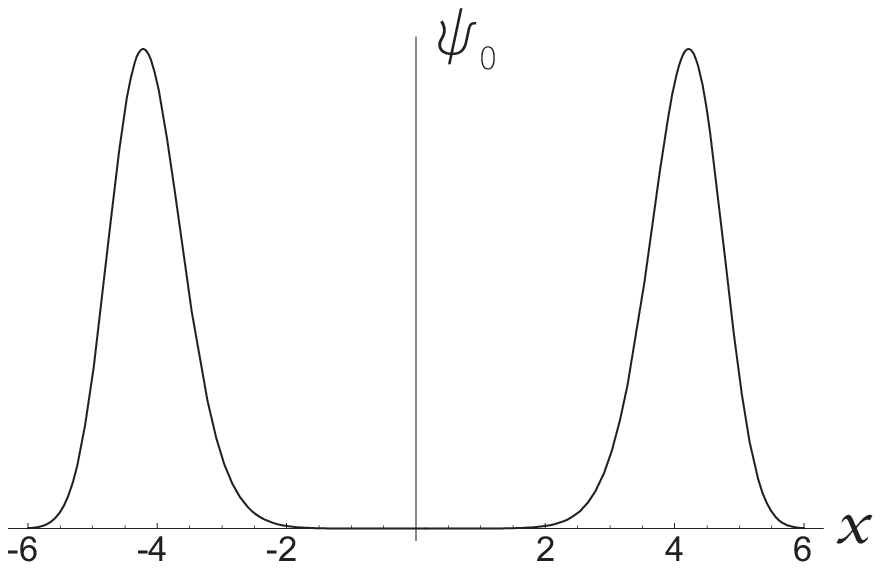,width=2.2in}\\ {\scriptsize{Ground state
$\psi_0(x)$}}
\end{tabular} &
\begin{tabular}{c}
\psfig{figure=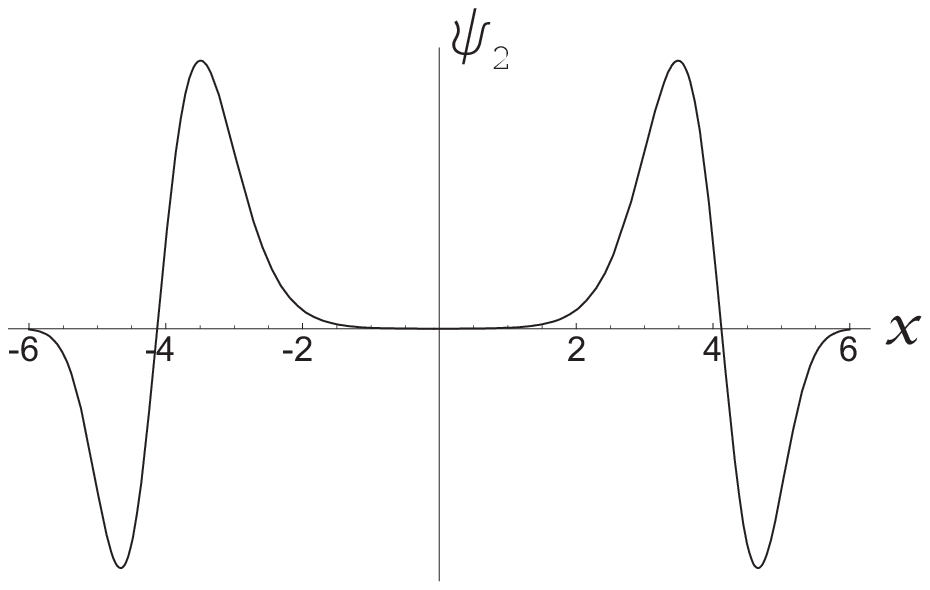,width=2.2in}\\ {\scriptsize{Second excited
state $\psi_2(x)$}}
\end{tabular}
\\\\
\begin{tabular}{c}
\psfig{figure=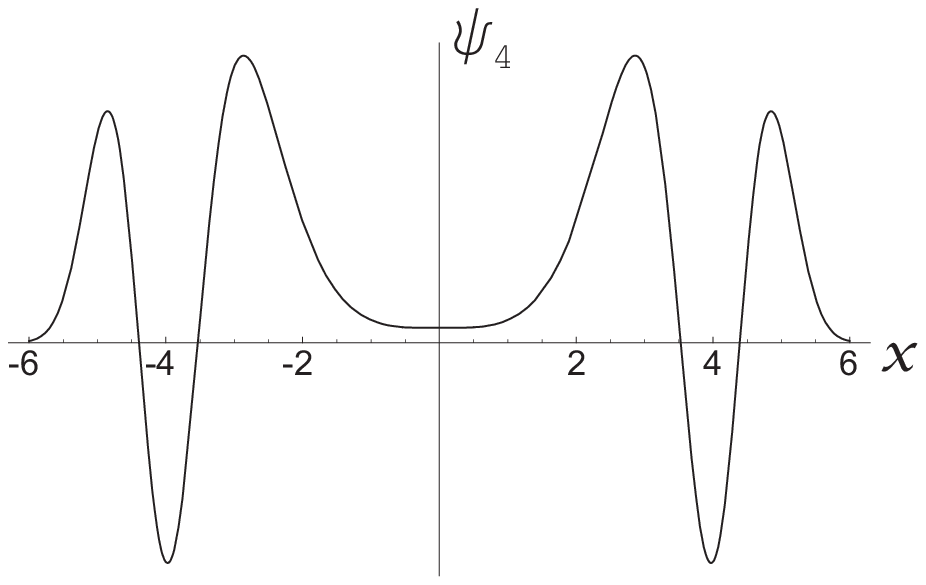,width=2.2in}\\
{\scriptsize{Fourth excited state $\psi_4(x)$}}
\end{tabular} &
\begin{tabular}{c}
\psfig{figure=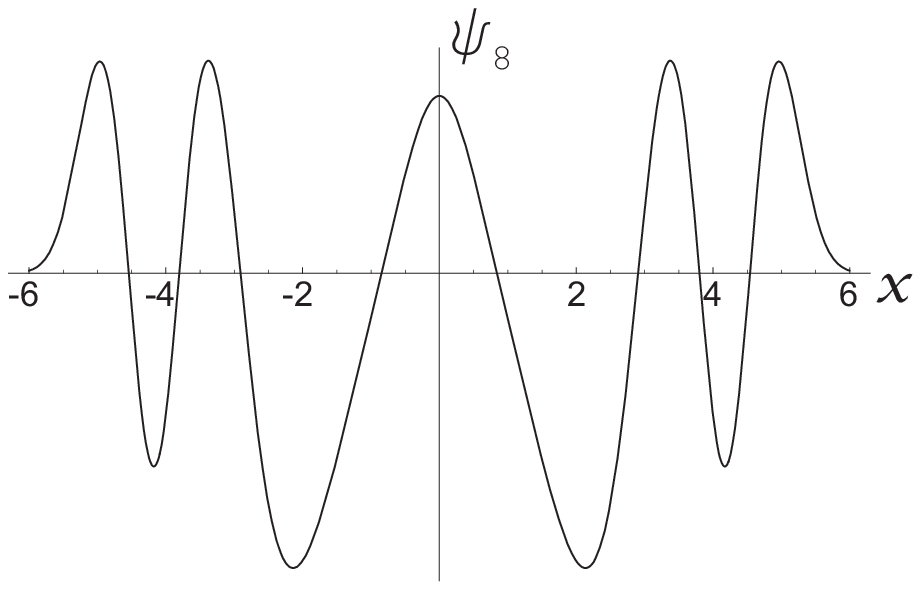,width=2.2in}\\ {\scriptsize{Eighth excited
state $\psi_6(x)$}}
\end{tabular}
\end{tabular}
\end{center}
\caption{Four even eigenfunctions of the modified sextic potential
(\ref{eq:sextic})  with $a=0.8$, $n=4$ and $p=0$, corresponding
to the exceptional module algebraization.}\label{fig3}
\end{figure}

\begin{figure}[h]
\begin{center}
\begin{tabular}{cc}
\begin{tabular}{c}
\psfig{figure=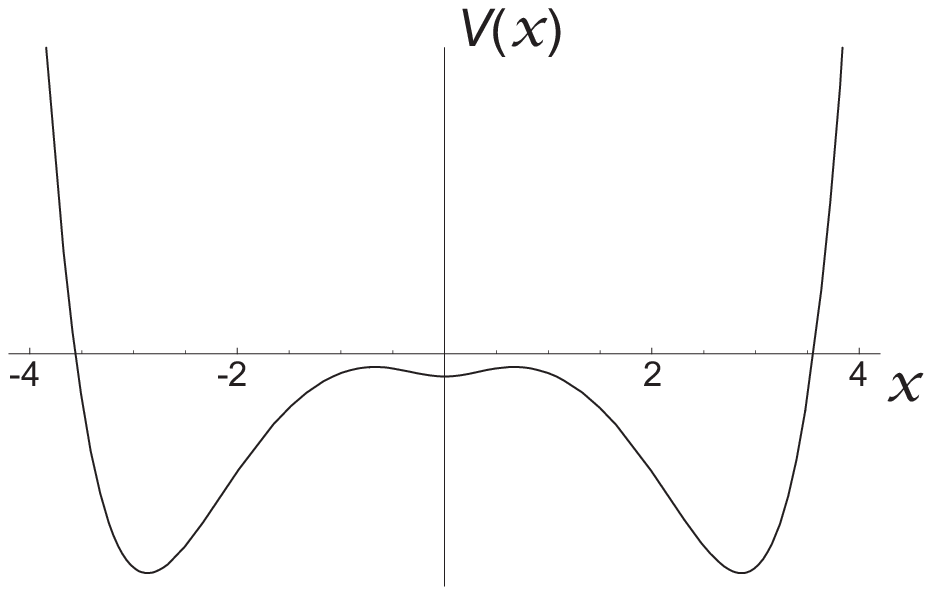,width=2.2in}\\{ \scriptsize{Potential
$V(x)$}}
\end{tabular} &
\begin{tabular}{c}
\psfig{figure=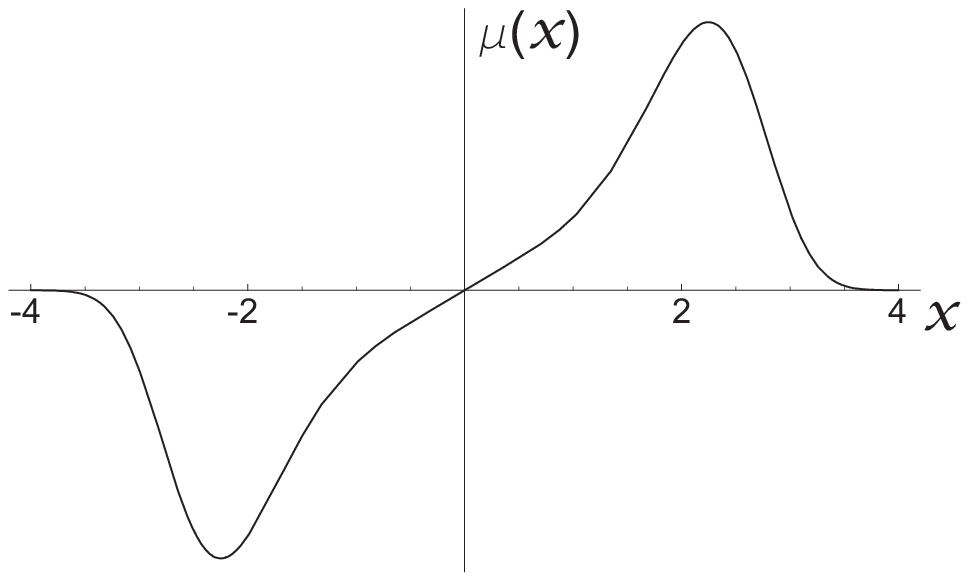,width=2.2in}\\ {\scriptsize{ Gauge factor
$\mu(x)$} }
\end{tabular}\\\\
\begin{tabular}{c}
\psfig{figure=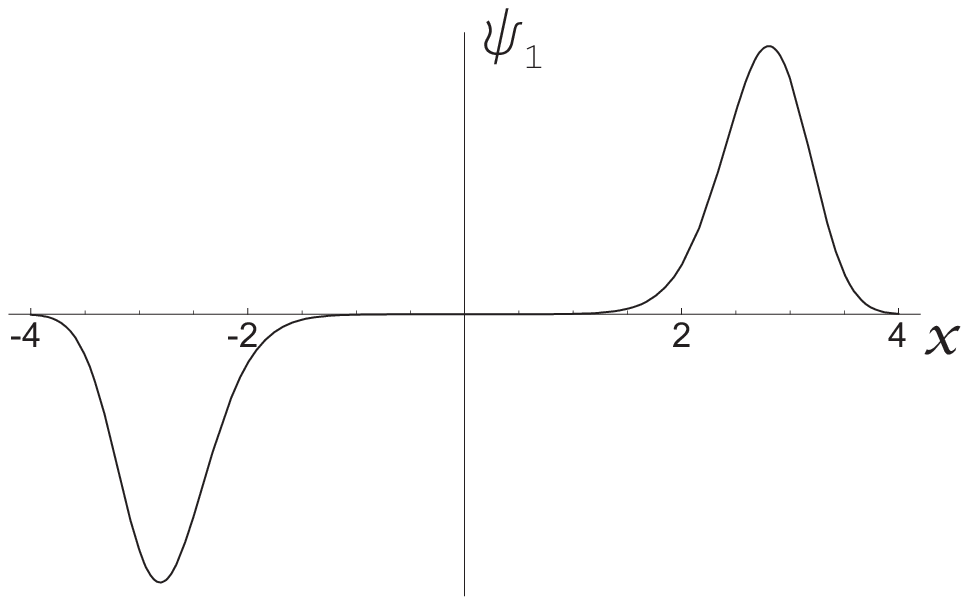,width=2.2in}\\ {\scriptsize{First excited
state $\psi_1(x)$}}
\end{tabular} &
\begin{tabular}{c}
\psfig{figure=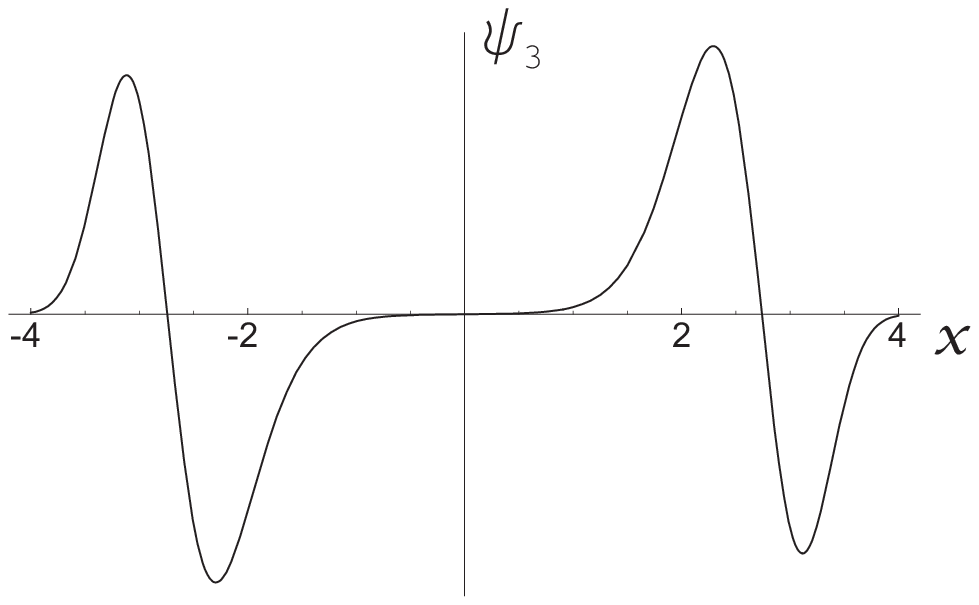,width=2.2in}\\ {\scriptsize{Third excited
state $\psi_3(x)$}}
\end{tabular}
\\\\
\begin{tabular}{c}
\psfig{figure=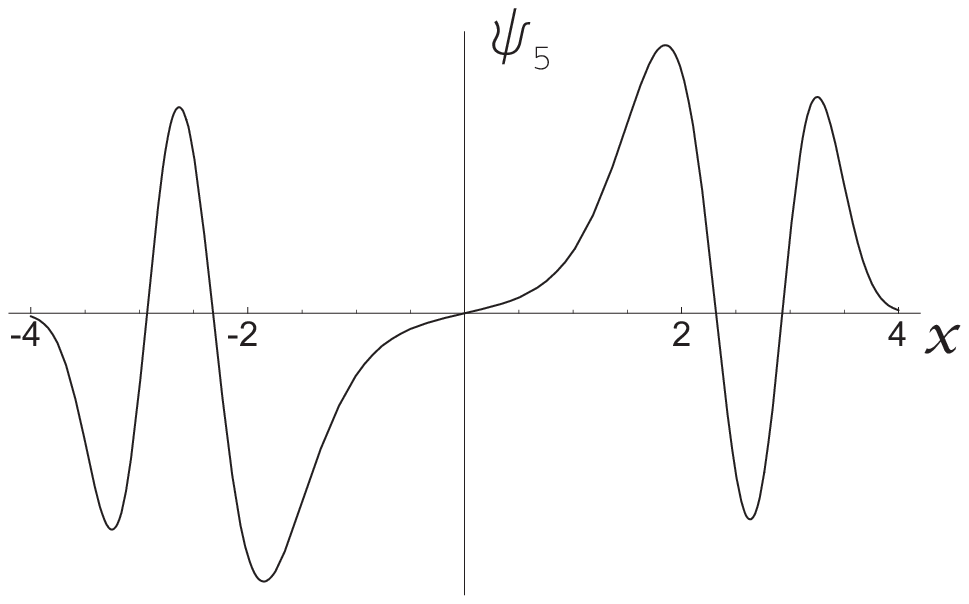,width=2.2in}\\
{\scriptsize{Fifth excited state $\psi_5(x)$}}
\end{tabular} &
\begin{tabular}{c}
\psfig{figure=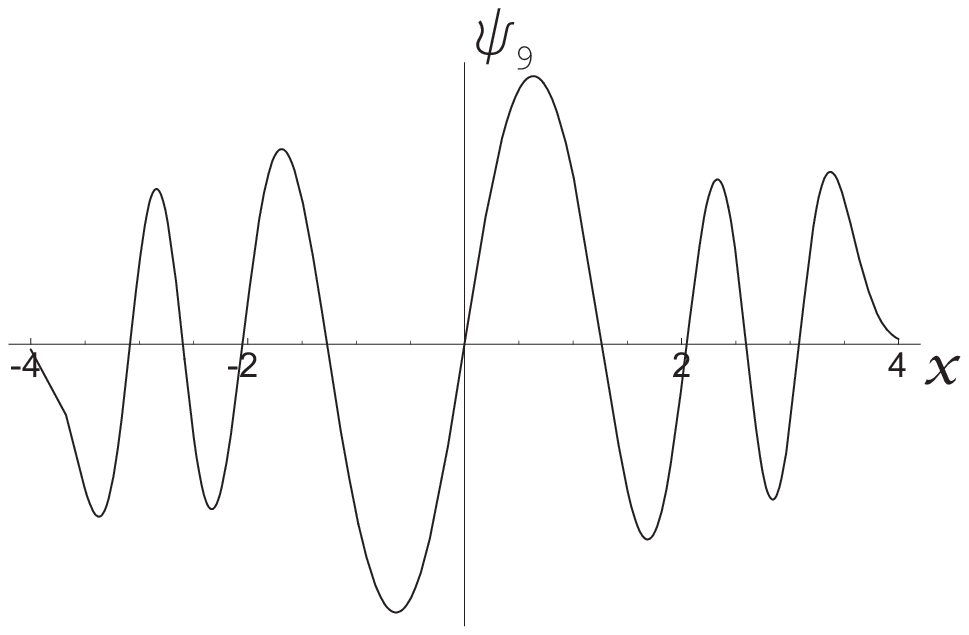,width=2.2in}\\ {\scriptsize{Ninth excited
state $\psi_9(x)$}}
\end{tabular}
\end{tabular}
\end{center}
\caption{Four odd eigenfunctions of the modified sextic potential
(\ref{eq:sextic})  with $a=0.8$, $n=4$ and $p=1$, corresponding
to the exceptional module algebraization.}\label{fig4}
\end{figure}

\noindent For $a=0.8$ the four polynomial eigenfunctions of $T$
are approximately
\begin{eqnarray*}
p_0(w)&=&0.431 +2.555 w + 3.543 w^2 +2.071 w^3 +0.325 w^4,\\
p_1(w)&=& 0.965+2.799w+2.681w^2+0.776w^3-0.055w^4,\\
p_2(w)&=&0.647+0.230 w+0.369w^2 -0.052 w^3+0.0017 w^4,\\
p_4(w)&=&0.329-0.538w+0.114w^2-0.008 w^3+0.0001 w^4,
\end{eqnarray*}where polynomial $p_k$ has $k$ zeros.
The corresponding energies are:
\begin{eqnarray*}
E_0&=&0,\\
E_2&=&6.058,\\
E_4&=&11.148,\\
E_8&=&15.129.
\end{eqnarray*}
Therefore the four even eigenfunctions of (\ref{eq:sextic}) coming
from the exceptional module algebraization are:
\begin{equation}
\psi_{2k}(x) = \mu(x)\,p_k(x^2),\qquad k=0,1,2,4.
\end{equation}
where the gauge factor $\mu(x)$ is given by (\ref{musextic}) with
$p=0$. For this choice of parameters $a$, $p$ and $n$ the
potential, gauge factor and eigenfunctions have been plotted in
Figure \ref{fig3}. We observe once again that we obtain
eigenfunctions with zero, two, four and eight zeros, but not one
with six zeros. For arbitrary $n$ the eigenfunction with $2n-2$
zeros would be missing form the algebraic sector. The calculations
for the odd case are similar: setting $p=1$ in (\ref{eq:sextic})
we have calculated $n=4$ odd eigenfunctions which are shown in
Figure \ref{fig4}.

\section{Discussion}\label{discussion}

By means of two examples we provide further arguments to motivate
the study of quasi-exactly solvable problems beyond the Lie
algebraic approach. Even in the simplest case of one dimensional
quantum mechanical problems the classifications performed in
\cite{T,GKO93} do not cover all quasi-exactly solvable potentials.
Since Lie algebraic potentials are only a subclass of the
potentials with partial algebraization of their spectrum, it would
be desirable to find a precise mathematical characterization of
the latter. This problem is related to the so called {\em
generalized Bochner problem} \cite{T94}, that of finding the most
general differential operator that preserves a general finite
dimensional space of polynomials. All the new quasi-exactly
solvable potentials obtained recently are based on the exceptional
modules, but other polynomial subspaces might exist which do not
have a monomial basis and yet have a rich structure. In
conclusion, the class of Hamiltonians with partial algebraization
of their spectrum is larger than it was previously thought, and
the new results call for new theoretical developments on this
field.

\section*{acknowledgements} The authors would like to thank
Prof. Pogosyan whose observations lead to some of the results of
this paper. The research of DGU is supported in part by the
Ram\'on y Cajal program of the Ministerio de Ciencia y
Tecnolog\'{i}a and by the DGI under grant FIS2005-00752. The
research of NK and RM is supported in part by the NSERC grants
RGPIN 105490-2004 and RGPIN-228057-2004, respectively.

\end{document}